# Direct evidence of non-ideal MHD in the solar wind and switchbacks

Forrest Mozer [1,2], Oleksiy Agapitov[1], Kyung-Eun Choi[1], Richard Sydora[3]
[1]Space Sciences Laboratory, University of California, Berkeley, 94720, USA.
[2]Physics Department, University of California, Berkeley, 94720, USA.
[3]Physics Department, University of Alberta, Edmonton T6G 2E1, Alberta, Canada.

The low frequency solar wind and its switchbacks are sometimes characterized as ideal-MHD plasmas containing Alfvenic turbulence and, sometimes, as non-ideal plasmas whose turbulence cannot be Alfvenic. The reason for this ambiguity is that definitive measurements of electric fields in the plasma rest frame at fluid frequencies have not been made and such measurements rule out the possibility of one or the other of the two modes. The FIELDS electric field measurement on the Parker Solar Probe is capable of such measurements, which are reported in this paper. At radial distances less than 40 solar radii on orbit 10, perpendicular and parallel electric fields in the plasma rest frame at frequencies of 0.01-0.1 Hz and over a broad frequency range are observed. The parallel electric fields survive in the presence of calibration errors because their amplitude, but not their existence, is affected by them. Errors in the perpendicular electric field or the solar wind velocity can produce large errors in the plasma rest frame perpendicular electric field because it is the difference of two large quantities, E and -vxB. Evidence is presented that this does not happen, at least much of the time. Because both perpendicular and parallel electric fields in the plasma rest frame are observed at low frequencies, the solar wind and its switchbacks reside in a non-ideal magnetohydrodynamic environment. Consequently, standard ideal-MHD treatments—including ideal Alfvén, slow, and fast wave formulations—cannot adequately describe the turbulent solar wind dynamics. Similar considerations may apply in other astrophysical plasmas.

## INTRODUCTION

Magnetic field, velocity, and density fluctuations in the ambient solar wind have been measured for many decades [Coleman 1968; Intriligator & Wolfe 1970] and the majority of our knowledge about solar wind turbulence has come from such measurements. This turbulence was originally thought to be predominantly Alfvenic [Belcher & Davis 1971] because the plasma was thought to be ideal. It has sometimes been treated as an ideal-MHD plasma at large fluid/inertial scales ($f \ll fci$) while invoking non-ideal Hall or kinetic physics at smaller kinetic scales or discrete reconnection boundaries [Sakshee et al, 2022; Bruno and Carbone ,2013; Verscharen et al, 2019].

The solar wind contains an additional component, magnetic switchbacks, whose prevalence and importance have been realized through analyses of data obtained by the Parker Solar Probe (PSP). Magnetic switchbacks are transient, large-amplitude, rotations of the interplanetary magnetic field [Raouafi et al, 2023; Badman et al, 2026; Wyper et al, 2026; Mozer et al, 2020a; Krasnoselskikh et al, 2020; Larosa et al, 2021; Dudok De Wit et al, 2020]. They are a class of disturbed magnetic fields in solar wind turbulence that might incorporate different processes than occur in the ambient medium. Switchbacks have been analyzed as Alfvenic processes in an ideal MHD environment [Landi et al, 2006; Squire et al, 2020; Tenerani et al, 2020; Ruffolo et al, 2020; Chen et al. 2020] and as products of a non-ideal plasma [Drake et al, 2021; Liang et al, 2021]. Early measurements [Mozer and Chen, 2013; Bale et al, 2005] suggested, through observation of electric fields in the plasma rest frame at low frequencies, that non-ideal MHD was at least sometimes present, but quantitative estimates of its importance have been lacking. A major reason for this situation is that, prior to PSP, useful electric field measurements in the solar wind and in switchbacks have been infrequent. Such data has become available from the FIELDS instrument [Bale et al, 2016] and it is used in this paper to obtain more detailed analyses of the solar wind turbulence. The electric field data in this paper is in the plasma rest frame and covers both the broad band of DC-150 Hz and 0.01-0.1 Hz, which is 2-3 orders-of-magnitude below the ion gyrofrequency of 3-9 Hz. Details of the electric field operation are given in Mozer et al, [2020b]. The two components of electric field in the plane perpendicular to the Sun-satellite line (the X-Y plane) were measured while the Z-component, which points Sunward, is assumed to be zero in plots of the total electric field. The solar wind velocity in this paper comes from the SWEAP/SPAN-I instrument on the Parker Solar Probe [Livi et al, 2022].

## IDEAL VERSUS NON-IDEAL MAGNETOHYDRODYNAMICS (MHD) IN SWITCHBACKS AND THE AMBIENT SOLAR WIND.

Ideal MHD assumes $\mathbf{E} + \mathbf{v} \times \mathbf{B} = \mathbf{E'} = 0$, where **E** is the electric field in the frame in which the plasma moves at the velocity **v,** and **E'** is the electric field in the plasma rest frame. These are the leading-order terms of the generalized Ohm's law for spatial scales much greater than the **i**on inertial length, (~6 km), and the ion gyroradius, (3 km), and frequencies much less than the ion cyclotron frequency, (~ 10 Hz). Because the Taylor scale (~15,000 km) of the present data at frequencies of 0.01-0.1 Hz is much larger, the electric field data of this paper is more than sufficient for deciding if or when the large-scale medium is in the ideal or non-ideal MHD state.

Figure 1 presents six days of Parker Solar Probe data on orbit 10 during which the spacecraft was below 40 solar radii from the Sun, as shown in Figure (1f). During the inbound orbit, the radial component of the magnetic field in Figure (1a) was large and variable while the total field of (1b) was essentially short-term constant, which signifies that the spacecraft crossed a high density of switchbacks. On the outbound leg, the radial component of the magnetic field was less variable, showing that the spacecraft passed through a more quiescent solar wind. In both regions, the magnitude of the plasma frame electric field (Figure 1c) was non-zero and as large as 300 mV/m. In this and all figures, the total electric field was obtained by assuming that the unmeasured $E_Z$' was zero.

The component of the electric field parallel to the background magnetic field is illustrated in Figure (1e). It was determined in the following way. The parallel electric field is given as

$$E_{\parallel} = (E_X B_X + E_Y B_Y + E_Z B_Z)/B \quad (1)$$

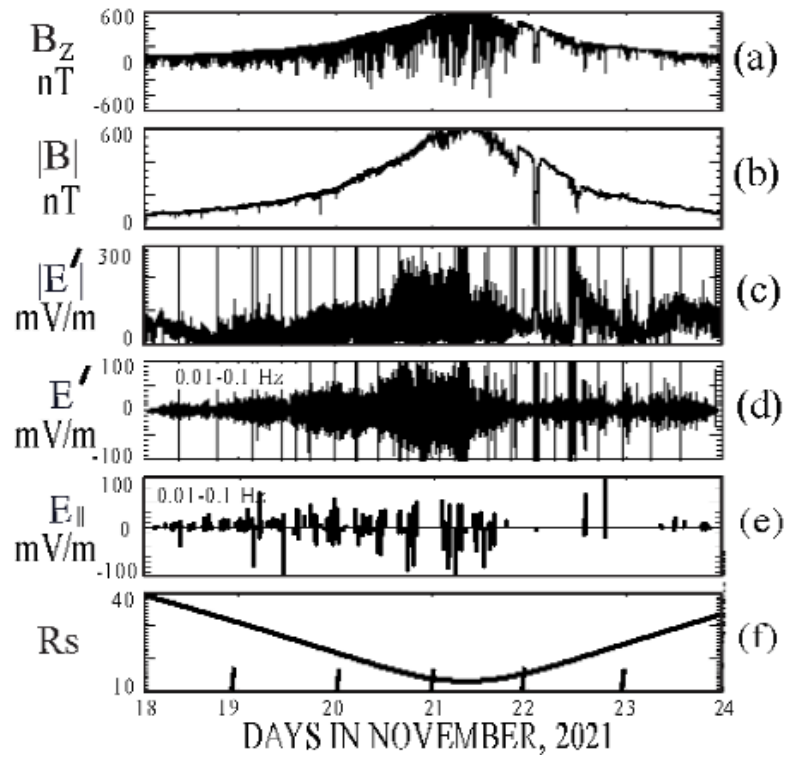

Figure 1. Six days of fields data on orbit 10 of the Parker Solar Probe. (1a): the radial (Z) component of the magnetic field, showing a high level of turbulence on the in-bound leg and less out-bound. (1b): the total magnetic field, whose short term near-constancy identifies the magnetic turbulence as switchbacks. (1c): The total electric field in the plasma rest frame, obtained by adding vxB to E and assuming that the unmeasured $E_Z$'was zero. (1d): the 0.01-0.1 Hz low frequency component of the plasma frame electric field. (1e): The 0.01-0.1 Hz electric field parallel to the magnetic field. (1f): the radial location of the spacecraft. Note that many of the vertical spikes in (1c) and (1d) result from on-board calibrations.

It is not generally measured on the Parker Solar Probe because the electric field component in the satellite-Sun direction, $E_Z$, was not measured. However, in switchbacks, the magnetic field sometimes rotated through more than 90°, at which time $B_Z$ became zero and $E_{\parallel}$ was determinable. The 0.01-0.1 Hz parallel electric field data of Figure (1e) is presented for all times when $|B_Z/B|<0.05$. Tests with $|B_Z/B|$ as large as 0.1 have shown a negligible leakage of the perpendicular electric field into the parallel field estimate. The parallel electric field may be present at times other than when $|B_Z/B|<0.05$, but, if so, it cannot be measured on the Parker Solar Probe.

Because both the perpendicular and parallel electric fields in the plasma frame at low frequencies were non-zero (1c), (1d), ideal-MHD cannot be used to determine any properties of the medium. Concepts such as Alfvenicity, slow mode and fast mode, which are associated with ideal-MHD, cannot apply to the large-scale solar wind.

The spatial scale of the described processes may be estimated from the Taylor hypothesis to be about the solar wind speed divided by 0.03 Hz or 15,000 km. The 0.01-0.1 Hz perpendicular field of 20-30 mV/m across 15,000 km corresponds to 450 kV, which seems large until it is compared with the convection field of 300 mV/m across the same distance, which is 4,500 kV. The perpendicular plasma frame potential is thus 10 percent of one that is present everywhere in the solar wind at all times, and is in no way unphysical. The 600 V is alarming only if applied to the parallel field, which is a separate matter treated below.

The spectra of the electric (2a) and magnetic (2b) fields on the full days of important switchback activity (November 21, 2021) and weak activity (November 23, 2021) are illustrated in Figure 2, with the black curves being the first of these two days and the red curves being the latter day. They illustrate the turbulent nature of the plasma. The waveform of the electric field over the frequency range of 0.01-0.1 Hz is illustrated in Figure (1d). This frequency range will be more carefully examined in the coming figures.

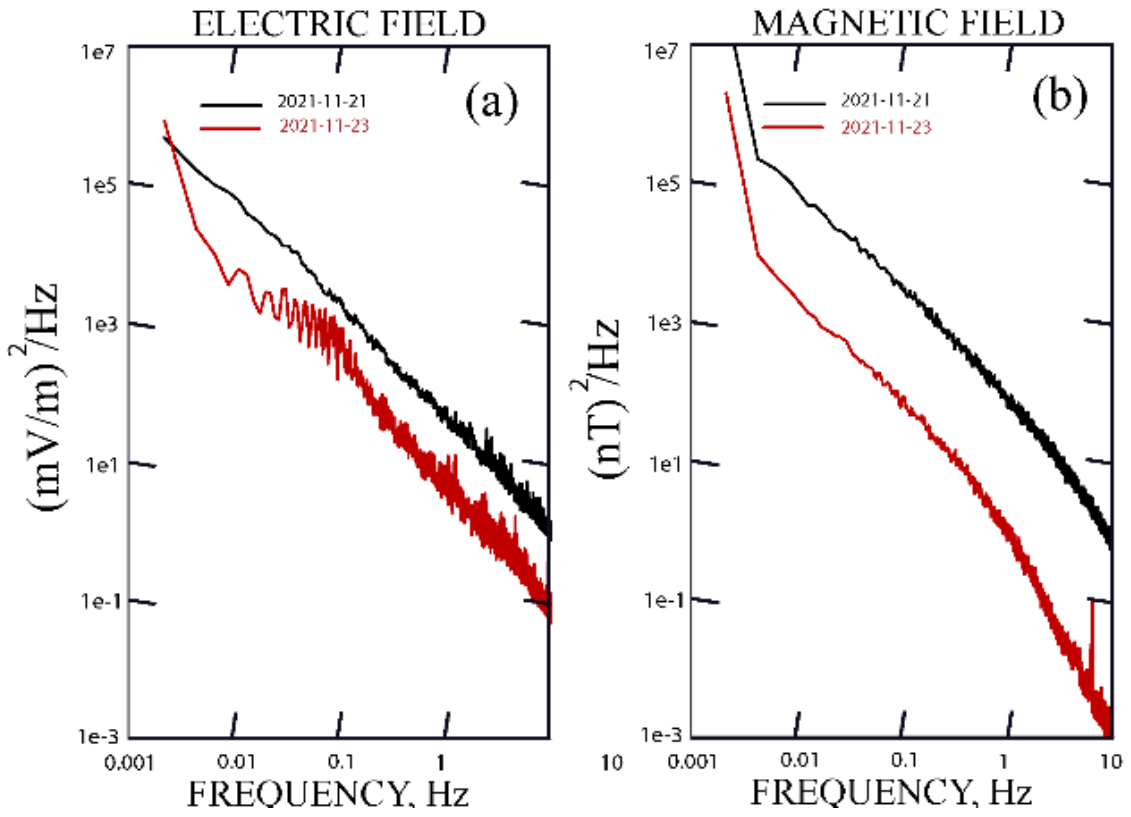

Figure 2. Spectra of the electric field (2a) and magnetic field (2b) during days of dense switchbacks and quiescent plasma.

PROPERTIES OF A SWITCHBACK

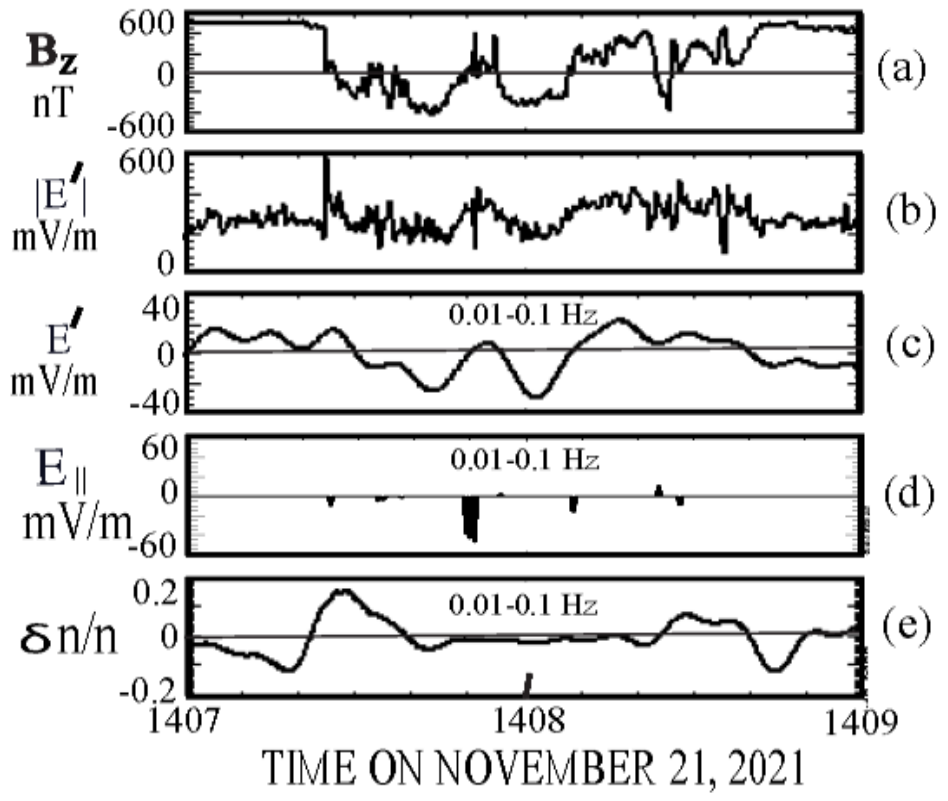

Figure 3. The fields for two minutes inside a switchback. (3a): the radial component of the magnetic field. (3b): the total electric field. (3c): the low frequency component of the electric field inside the switchback. (3d): the 0.01-0.1 Hz parallel electric field. (3e): low frequency (0.01-0.1 Hz) density fluctuations obtained from the SWEAP data.

Figure (3a) describes a 140° rotation switchback having an electric field as large as 400 mV/m (3b) and an accompanying low frequency electric field (Figure 3c) as large as 30 mV/m. Low frequency parallel electric fields (3d) as large as -40 mV/m were also observed. Low frequency density fluctuations larger than 10% of the total density (3e) were present. These show that the plasma was compressive.

The observed intense electric fields must be balanced by terms in the Generalized Ohm's Law, whose leading terms are

$$\mathbf{E} + \mathbf{v} \times \mathbf{B} = \eta\, \mathbf{J} + (\mathbf{J}\times\mathbf{B}) /(ne) - (\nabla\cdot\mathbf{P}_e)/(n\, e) + [m_e/(ne^2)]\, [\partial\mathbf{J}/\partial t + \nabla\cdot(\mathbf{vJ} + \mathbf{Jv})] \qquad (2)$$

where

$\mathbf{E} + \mathbf{v}\times\mathbf{B}$ is the electric field in the plasma rest frame.

$\eta\mathbf{J}$ is the resistivity from collisions or anomalous wave-particle scattering.

$\mathbf{J}\times\mathbf{B}/(ne)$ is the Hall term.

J is the current density in the plasma

n is the plasma density and

e is the charge on an electron

$-\nabla\cdot\mathbf{P}_e/(ne)$ is the electron pressure divergence.

$[m_e/(ne^2)][\frac{\partial \mathbf{J}}{\partial t}+\ldots]$ is the electron inertia.

Analyses of the magnitudes of these terms in the solar wind [Shoda et al, 2021; Priest and Forbes, 2000], application of these terms to earlier data [Verscharen et al, 2019; Bale et al, 2005] and simulations or theories [Sahraoui et al, 2009; Schekochihin et al, 2009; Howes et al, 2008] have shown that the Hall term dominates at the scales of interest. Thus, for the perpendicular electric field, we assume:

$$\mathbf{E}'_{\perp} = \mathbf{J}\times\mathbf{B}/ne \qquad (3)$$

Because only $E'_X$ and $E'_Y$ were measured, this equation may only be directly solved for the Z-component of the current density, which reached 50 microamps/$m^2$. A current of this magnitude cannot be distributed across the structure as a whole. By Ampère's law a 50 nT field change carries it over 0.8 km and a 500 nT change over 8 km, so the Hall current must be concentrated in layers of order the ion inertial length embedded within the much larger rotation, which is consistent with the ion decoupling. The low ion plasma beta (<0.5) in this event indicates that ions were inertially decoupled. In this regime, the ions could not track the rapid magnetic fluctuations, while the magnetized electrons carried the current, generating the observed Hall electric field.

The current must close, since $\nabla\cdot\mathbf{J} = 0$. Within the switchback itself no closure is required because the current derived from equation (3) is perpendicular to the direction of convection by construction. The divergence of the perpendicular current must be balanced by a field-aligned current carried along the magnetic field by the electrons. The demand that this places on the plasma is slight. Closing a perpendicular current of 10 μA $m^{-2}$ requires a field-aligned electron drift of ≈ 60 km $s^{-1}$, which is ~1 percent of the electron thermal current. The closure current is therefore available at negligible cost.

The parallel electric field must also be balanced by terms in the Generalized Ohm's Law, not including the Hall term because that term is perpendicular, not parallel, to the magnetic field. The three terms, electron pressure gradient, electron inertia, and an effective resistivity may be distinguished by the scale each would require to produce the measured field. For $E_{\parallel}{\sim}40$ mV/m, an electron temperature ~50 eV, and a density ~$10^3$ $cm^{-3}$, the electron pressure gradient is the only term that can supply a field of the observed size, and it requires gradients on scales of order 100 meters to 1 kilometer, which is somewhat greater than the electron inertial length. Because the Hall term does no work, $E_{\parallel}$ is the only term that can heat the plasma.

VALIDITY OF THE ELECTRIC FIELD MEASUREMENTS

The parallel electric field measurements of Figures (1e) and (3d) may be in error due to using an inaccurate calibration of the instrument. However, this could only affect their amplitude, not their existence, so such data presents solid evidence of the non-ideal nature of the MHD.

Perpendicular electric field measurements might be perturbed by wakes around the spacecraft or electronic asymmetries or malfunctions, etc. To show that none of these problems occurred and that the electric field data is of the highest quality, measurements of the potential differences between each of the four sensors and the spacecraft body are analyzed in Figure 4. Because the sensors responded to both the nearby plasma density and the external electric field, the two effects must be separated. The average of the potentials of the four sensors responds to plasma density changes, so it is subtracted from each sphere potential and converted into the density estimate illustrated in (4a). The relationship between the spacecraft potential and the plasma density results from the requirement that the outflowing high energy end of the dominant photoemission current (which is proportional to exp(-Vsc)) being equal to the incoming plasma current (which is approximated as the plasma density times the square root of the electron temperature). Thus,

$$Vsc = A*\ln(nT^{0.5}) + B$$

where

Vsc is the spacecraft potential found as the average of the four sensor potentials
n is the plasma density and T is the electron temperature
A and B are least-squares coefficients

The agreement between the two density estimates in (4a) is satisfactory given the neglect of secondary emission, spectral changes, etc. in the current balance. This result provides confidence that the electric field sensors did not encounter wakes, mechanical or electrical offsets, etc.

The remaining components of the sensor potentials (after removing their average values) are illustrated in (4b) and (4c). That the potentials of opposite pairs of sensors anti-correlate with high accuracy shows that the sensors responded to an external electric field rather than to spacecraft wakes, local density structures, or electronic asymmetry, and that no malfunction was present. Note that the electric field is formed from (V2-V1) and (V4-V3).

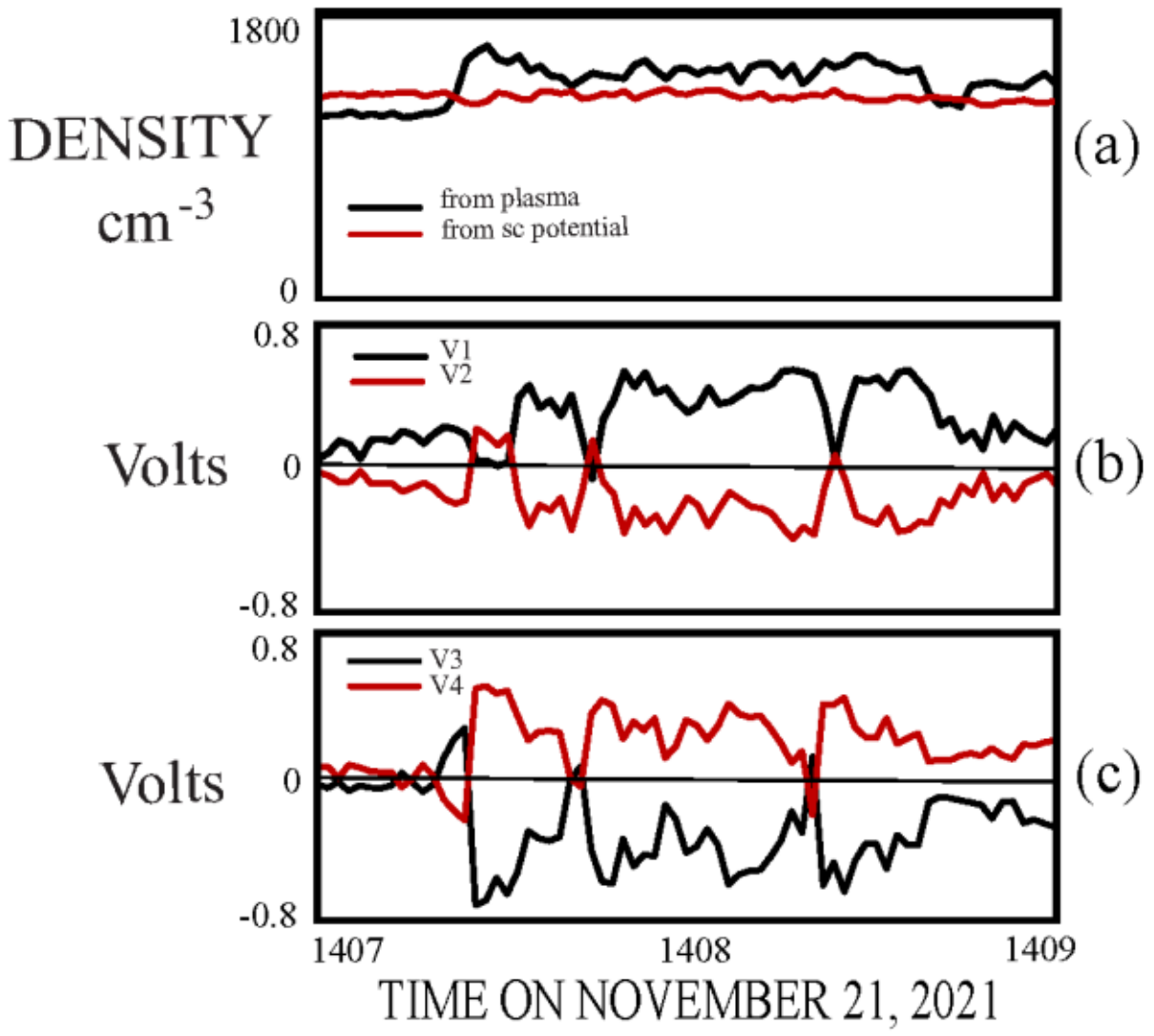

Figure 4. The four spacecraft potentials broken into two parts. The average of the four potentials plus two least squares coefficients produce the plasma density estimate of (4a) along with that obtained from the plasma data. The remaining four potentials are illustrated in (4b) and (4c). The anti-correlation between V1 and V2 and between V3 and V4 shows that the experiment worked inside the switchback with no wakes, electronic malfunctions, etc.

The perpendicular electric field measurements result from the small difference of two large quantities. Thus, a relatively small error in the solar wind velocity or the electric field could result in a large error in the electric field in the plasma rest frame. This problem is most serious inside switchbacks because the magnetic field has a large component perpendicular to the radial velocity, so vxB is large. This situation is illustrated in Figure 5, which describes the rest frame Poynting flux (5a) inside the switchback of Figure (3a) with the spacecraft frame electric field components (the black curves of Figures (5b) and (5c)) and the components of -vxB (the red curves). Because the electric field and -vxB differed by ~10-20% in this switchback, relatively small errors in either measurement would lead to an invalid estimate of the rest frame electric field. That this does not happen is illustrated in Figure (5a), where the Poynting flux inside the switchback has a value consistent with expectations while that outside, at the beginning and end of the plot, the Poynting flux is close to zero as expected in the solar wind. It would be difficult for a non-valid measurement to produce these results by luck.

To further investigate the relationship between E in the spacecraft frame and -vxB, the two quantities are plotted for two days in Figures (6b) and (6c). Because the electric fields were much larger than -vxB, especially in (6b), measurement errors of either kind might modify the magnitude of the electric field in the plasma frame, but not its existence.

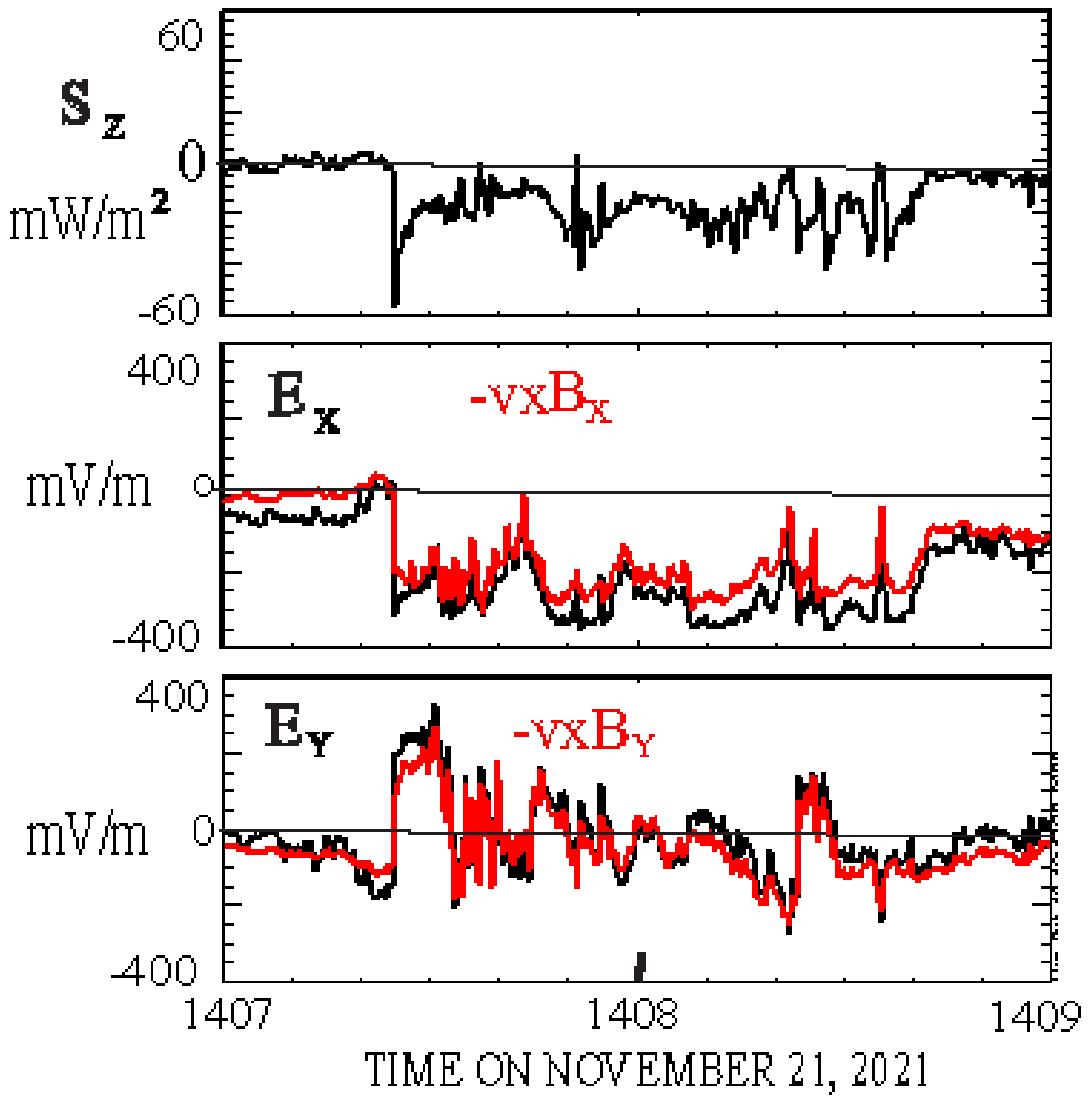

Figure 5. (5a): the plasma rest frame Poynting flux through and on each side of the switchback. (5b) the X-component of the electric field and vxB. (5c) the Y-component of the electric field and vxB. Because E and -vxB are similar in magnitude, a calibration error in either would cause their difference to become large. That this does not happen is evidence by the Poynting flux having a non-zero, reasonable value inside the switchback and being small outside, where it should be zero.

Figure 6 describes a day of plasma frame, 0.01-0.1 Hz electric fields and -vxB, (6b) and (6c), as well as the low frequency parallel electric field (6d) observed when $|B_Z|$ was nearly zero. Because the electric field exceeded -vxB by a large factor most of the time, calibration or offset errors cannot explain the non-zero value of E' observed for a large part of the day. Thus, both the low frequency parallel electric field and the rest frame low frequency perpendicular electric field exist to show that the measurements were well-made and that the plasma cannot be described by ideal-MHD during this time.

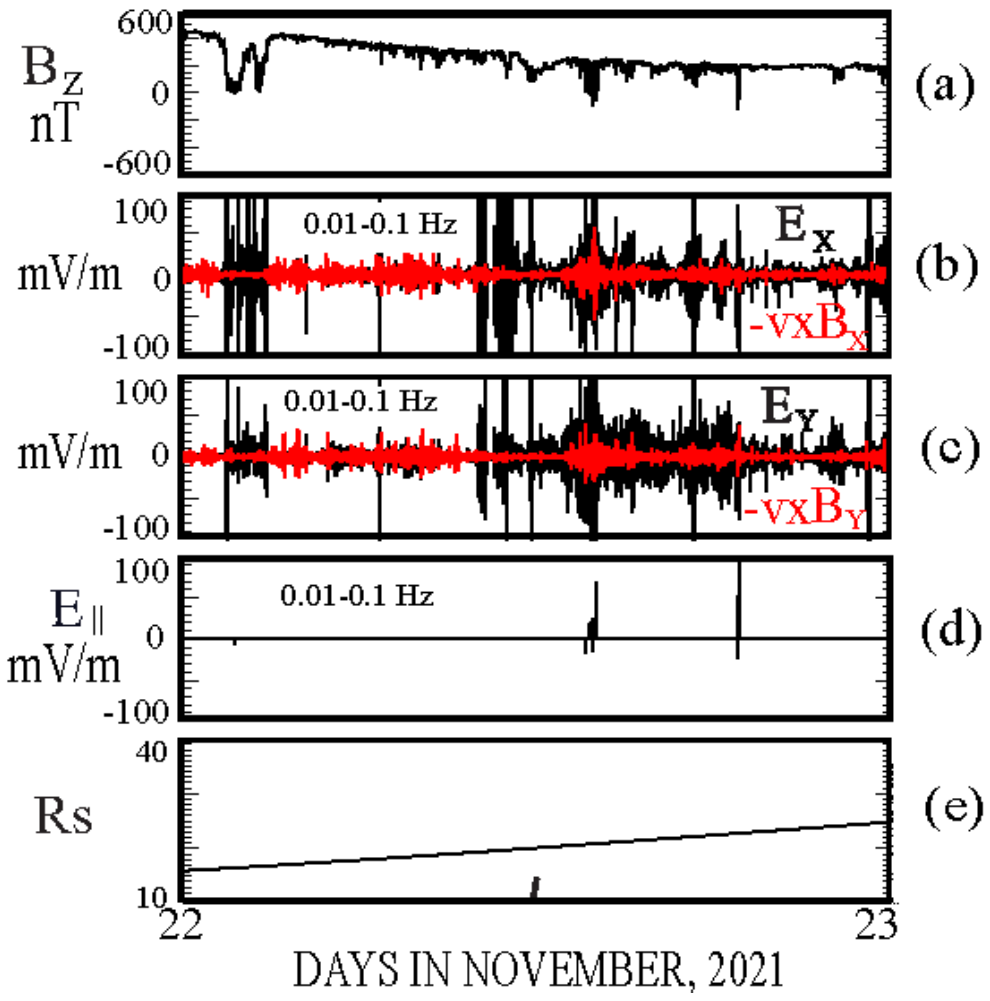

Figure 6. A day of magnetic field (6a), broadband 0.01-0.1 Hz electric fields in the spacecraft frame (6b) and (6c) along with the appropriate components of the low frequency -vxB. (6d): the 0.01-0.1 Hz parallel electric field. (6e): the spacecraft distance from the Sun. Because E during these days was much larger than -vxB, a calibration error in either might change their sum and the low frequency plasma frame electric field magnitude, but not its existence.

## EFFECT OF THE PARALLEL ELECTRIC FIELD ON THE PLASMA

The generalized Ohm's law is just the electron momentum equation. Setting $E_\parallel \approx -(\nabla\cdot P_e)_\parallel/(ne)$ says that the bulk electron velocity does not change. But the pressure term is not a force acting on any particle. It is the net momentum flux carried by electrons streaming in and out of a volume, so it is a kinetic, collective effect. The only forces on a single electron are the parallel electric field and the mirror force. Each electron moving along **B** therefore gains or loses parallel energy. Ohm's law says that these per-particle energy changes, summed over the distribution, are matched by the density and pressure changes along the path. The fluid stays in a steady state while every particle is accelerated or decelerated.

In short, Ohm's law balance is a statement about the first moment of the distribution. It limits bulk electron acceleration, not the acceleration of individual electrons, which would show up as energy shifts in the distribution and as net ion acceleration. An effort is underway to investigate such shifts.

## CONCLUSIONS

Because the Taylor scale (~15,000 km) of the present data at frequencies of 0.01-0.1 Hz is much greater than the ion inertial length, (~5 km), and the ion gyroradius, (~5 km), and the frequencies are much less than the ion cyclotron frequency, (~10 Hz), the electric field data of this paper is more than sufficient for determining if or when the medium is in the ideal or non-ideal MHD state. Because both the parallel and the perpendicular electric fields in the plasma frame are non-zero in switchbacks and in the quiescent solar wind at frequencies of 0.01-0.1 Hz and over a broad frequency band, the fluid plasma cannot be in the ideal-MHD plasma state. Thus, concepts like

Alfvenicity, slow mode waves, and fast mode waves, cannot be applied to the solar wind or its switchbacks at large scales. Models relying on ideal Walén relations fail to capture the underlying physics, requiring a transition to Hall-MHD or kinetic wave-particle frameworks.

The 15,000 km scale is the distance traversed by the spacecraft, not a coherence length, and within a switchback it lies largely across the magnetic field rather than along it. It is therefore not an integration path for the parallel field. The currents and parallel field gradients inferred above occupy thin layers embedded within that much larger structure; the two scales coexist and are not alternatives.

ACKNOWLEDGEMENTS

We thank the NASA Parker Solar Probe team and the FIELDS instrument suite principal investigators. The instruments were developed under NASA contract NNN06AA01C.

**Data Availability Statement**

The data used in this study were collected by the NASA Parker Solar Probe (PSP) mission. The Level 2 and Level 3 data from the FIELDS and SWEAP instrument suites for Orbit 10 are publicly available via the Space Physics Data Facility (SPDF) Coordinated Data Analysis Web (CDAWeb) at https://cdaweb.gsfc.nasa.gov/. High-resolution FIELDS data and processing routines used for the calculation of Poynting flux and wave-particle interaction metrics can also be accessed through the Berkeley Space Sciences Laboratory (SSL) Parker Solar Probe Gateway at https://fields.ssl.berkeley.edu/data/. Source data for the figures presented in this manuscript are available from the corresponding author upon reasonable request.